\documentclass[a4paper,11pt]{article}
\usepackage{pos}

\usepackage{natbib,natbibspacing}
\usepackage[export]{adjustbox}
\DeclareUnicodeCharacter{2212}{-}

\newcommand{\hess}{H.E.S.S.}
\newcommand{\at}{AT2019krl}

\title{\hess\ ToO program on nearby core-collapse Supernovae: search for very-high energy $\gamma$-ray emission towards the SN candidate AT2019krl in M74}
 \ShortTitle{\hess\ observations of core-collapse SNe and AT2019krl}

\author*[a]{Nukri Komin}
\author[b]{Matthieu Renaud}
\author[c]{Rachel Simoni}

\author[d]{Stuart Ryder}

\affiliation[a]{School of Physics, Wits University, Johannesburg}
\affiliation[b]{LUPM/CNRS, Montpellier, France}
\affiliation[c]{University of Amsterdam, Netherlands}

\affiliation[d]{Macquarie University, Sydney, Australia}

\forColl{\hess} 

\emailAdd{contact.hess@hess-experiment.eu}
\emailAdd{nukri.komin@wits.ac.za}
\emailAdd{r.c.simoni@uva.nl}
\emailAdd{matthieu.renaud@umontpellier.fr}

\abstract{While the youngest known supernova remnants, such as Cassiopeia A, have been proven to be able to accelerate cosmic rays only up to $\sim$10$^{14}\,\mathrm{eV}$ at their present evolutionary stages, recent studies have shown that particle energies larger than a few PeV ($10^{15}\,\mathrm{eV}$) could be reached during the early stages of a core-collapse Supernova, when the high-velocity forward shock expands into the dense circumstellar medium shaped by the stellar progenitor wind. Such environments, in particular the type IIn SNe whose progenitors may exhibit mass-loss rates as high as $10^{−2}M_\odot\,\mathrm{yr}^{−1}$ \cite{smith14}, could thus lead to $\gamma$-ray emission from $\pi^0$ decay in hadronic interactions, potentially detectable with current Cherenkov telescopes at very-high energies. Such a detection would provide direct evidence for efficient acceleration of CR protons/nuclei in supernovae, and hence new insights on the long-standing issue of the origin of Galactic Cosmic Rays. In that context, the High Energy Stereoscopic System (\hess) has been carrying out a Target of Opportunity program since 2016 to search for such an early very-high-energy $\gamma$-ray emission towards nearby core-collapse supernovae and supernova candidates (up to $\sim 10~\mathrm{Mpc}$), within a few weeks after discovery. After giving an overview of this \hess\ Target of Opportunity program, we present the results obtained from the July 2019 observations towards the transient \at, originally classified as a type IIn supernova, which occurred in the galaxy M74 at $\sim 9.8\,\mathrm{Mpc}$. Although its nature still remains unclear, the derived \hess\ constraints on this transient are placed in the general context of the expected VHE $\gamma$-ray emission from core-collapse supernovae.}

\FullConference{37$^{\rm{th}}$ International Cosmic Ray Conference (ICRC 2021)\\
		July 12th -- 23rd, 2021\\
		Online -- Berlin, Germany}

\begin{document}
\maketitle

\section{Introduction: probing particle acceleration in core-collapse supernovae}
\label{Intro}
Among the $\gamma$-ray emitting supernova remnants (SNRs), no indisputable evidence for the presence of PeV  Cosmic Ray (CR) particles being accelerated at their shock fronts has been found yet. Cassiopeia A (Cas~A), with an age of about 350 years one of the youngest Galactic SNRs, has long been considered as the best candidate to accelerate particles up to the CR knee ($\sim$3 PeV). Its high-energy/very-high-energy (HE/VHE) $\gamma$-ray spectrum measured with {\it Fermi}-LAT \cite{yuan13,ahnen17}, VERITAS \cite{kumar15} and MAGIC \cite{ahnen17} is better explained by hadronic interactions owing to the turnover seen below 1 GeV, reminiscent of the characteristic $\pi^0$ decay signature. The latest combined {\it Fermi}-LAT/VERITAS spectrum clearly shows an exponential cutoff of the $\gamma$-ray spectrum at $\sim$2-3 TeV \cite{veritas20}. This demonstrates that the maximum energy of the accelerated particles is below 20 TeV, regardless of the details of the  broadband spectral modeling of Cas~A \cite{veritas20}. Such a low value points towards the idea that PeV CRs could be accelerated at the shock front much earlier than previously thought during the SNR evolution. Recent studies of the diffusive shock acceleration (DSA) mechanism have shown that particle energies larger than a few PeV could actually be reached when the supernova (SN) shock is propagating at high velocity ($v_{\rm sh} \gtrsim 10^4$ km s$^{-1}$) into a dense ($n \gtrsim 10^{7-9}$ cm$^{-3}$) circumstellar medium (CSM) \cite{tatischeff09,bell13,schure13,murase14,cardillo15,bykov18,marcowith18}. Such an environment is formed by the wind from the stellar progenitor prior to explosion, whose density profile is commonly assumed to be steady: $n(R_{\rm sh}) = \frac{\dot{M}}{4 \pi u_{\rm w} \mu m_H R^2_{\rm sh}}~\sim~3\times{10^{7}}~\frac{\dot{M}_{-5}}{u_{\rm w,10}}~R^{-2}_{\rm sh,15}~~{\rm cm}^{-3}$ (for a mean mass per particle $\mu$ = 1) with a mass-loss rate $\dot{M} = 10^{-5} \dot{M}_{-5} M_{\odot}$ yr$^{-1}$, a wind velocity $u_{\rm w} = 10 u_{\rm w,10}$ km s$^{-1}$ and a shock radius $R_{\rm sh} = 10^{15} R_{\rm sh,15}$ cm. Tatischeff \cite{tatischeff09} has shown that the extensive spectro-morphological radio observations of the synchrotron emission from the nearby, and well-monitored, type IIb SN~1993J can be described by a detailed model coupling the non-linear DSA with self-similar solutions for the SN hydrodynamics. The magnetic field was found to be strongly amplified shortly after the explosion at the level of $\sim50~(t/1 {\rm day})^{-1}$ G\footnote{at least two orders of magnitude higher than the equipartition value of a Red SuperGiant (RSG) progenitor wind, in line with several previous studies \cite{fransson98,marti-vidal11}.}, most likely due to the CR resonant and nonresonant instabilities operating in the shock precursor (but see \cite{bjornsson17} for a different interpretation). Protons and nuclei would then be accelerated to PeV energies within the first few days after the explosion (see \cite{marcowith18}), with a total CR energy $E_{\rm CR} \sim 7 \times 10^{49}$ erg over the first $\sim$8 yr. From the best-fit values of the relevant parameters obtained by \cite{tatischeff09}, the hadronic $\gamma$-ray flux from SN~1993J amounts to 
\begin{equation}
F_\gamma(E>E_{\rm TeV})~\sim~2~\times{10^{-12}}~E_{{\rm TeV}}^{-1}~d_{{\rm Mpc}}^{-2}~\left(\frac{\dot{M}_{-5}}{u_{\rm w,10}}\right)^2~t_{\mathrm{day}}^{-1}~~{\rm cm}^{-2}~{\rm s}^{-1}.
\label{eq:1}
\end{equation}
Similar calculations have been performed \cite{dwarkadas13,murase14,zirakashvili16,fang19} in the case of the superluminous type IIn SNe, thought to be powered by the interaction between the ejecta and a very dense CSM \cite{gal-yam12}. Once the collisionless shock propagates in the wind, efficient CR acceleration has been shown to take place \cite{katz11,murase11}, and these SN events could thus be privileged source candidates to explain (part of) the very-high-energy neutrinos detected with IceCube \cite{zirakashvili16}. The expected $\gamma$-ray flux derived in \cite{murase14}, once applied on SN~1993J, is compatible with equation \ref{eq:1} within a factor of $\lesssim$ 2.


In the GeV domain, $\gamma$-rays are expected to be mostly unaffected by photon fields from the source \cite{murase14}, so that these flux estimates can be readily used to estimate the visibility of these SNe with current HE instruments such as {\it AGILE} and {\it Fermi}-LAT. The analysis of {\it Fermi}-LAT data towards an ensemble of 147 type IIn SNe in different time windows has not revealed any significant excess \cite{ackermann15}. Only for the brightest and closest type IIn events ($d \lesssim 20$~Mpc) do the derived upper limits start to constrain the theoretical expectations \cite{rt18}. Nevertheless, two recent studies based on {\it Fermi}-LAT data have reported marginally significant variable HE emission towards the peculiar H-rich super-luminous SN iPTF14hls at $\sim$150 Mpc \cite{yuan18} and the nearby ($\sim$3.5 Mpc) type IIP SN~2004dj \cite{xi20}. The latest search for $\gamma$-rays from SNe by means of a variable-size sliding-time-window analysis of the {\it Fermi}-LAT data has confirmed the variable signal in the direction of iPTF14hls and revealed two new excesses in the direction of the SN candidates AT2019bvr and AT2018iwp, with a flux increase within six months after the discovery date \cite{prokhorov21}. In the TeV domain, the pair production process $\gamma + \gamma \rightarrow e^+ + e^-$ in the radiation field from the SN photosphere is expected to be effective. While Tatischeff \cite{tatischeff09} has overestimated the opacity $\tau_{\gamma \gamma}$ for SN~1993J by assuming an {\it isotropic} source of UV/optical photons, the full opacity calculations, including temporal and geometrical effects due to the {\it anisotropic} $\gamma$-$\gamma$ interaction, have recently been applied to SN~1993J \cite{cristofari20}. The resulting $\gamma$-ray flux, shown to be sensitive to the photospheric properties and SN hydrodynamics, is then strongly attenuated during the first $\sim$ 10 days.

One should note that all these calculations are subject to large uncertainties in the different parameters highlighted above, as most of them are estimated through indirect means, once multi-wavelength (MWL) observations have been carried out. First of all, many details impacting the particle acceleration capability of these SNe are still poorly constrained (e.g. the magnetic field geometry, the nature and growth timescale of the instabilities and the particle injection). Moreover, the characteristics of cc-SN light-curves and spectra, the properties of the stellar progenitors (impacting the time evolution of the photospheric parameters $R_{\rm sh}$, $u_{\rm sh}$) and their stellar winds (mass-loss rate, wind velocity and structure\footnote{Recent observations show that some type Ibn/IIn SN progenitors (possibly Luminous Blue Variables) experience some episodic mass-loss outbursts during the latest phases before explosion, resulting in a complicated CSM \cite{smith17} made of shell-like structures \cite{nyholm17}.}) are very diverse and are still the topic of intense investigation \cite{smartt09b,smith14}. Also, the time of the SN's explosion is usually not known at the time of its discovery until detailed analyses of pre- and post-explosion data are performed. Nevertheless, high-cadence UV/Optical/NIR observations\footnote{ASAS-SN {\scriptsize \url{(http://www.astronomy.ohio-state.edu/~assassin/)}} and DLT40 {\scriptsize \url{(http://dark.physics.ucdavis.edu/dlt40/DLT40)}} projects are yielding the cadence necessary to ensure that nearby southern cc-SNe can be caught within $<$3 days after the explosion.} allow for a determination of the SN subclass within a few days and provide important insights on the SN photosphere, the stellar progenitor and on the ejecta chemistry and kinematics (e.g. \cite{nagao19,bose20}). Finally, radio observations provide a probe of the magnetic fields in core-collapse supernovae (cc-SNe), as well as the mass-loss rate and the density distribution of the CSM produced by the pre-SN stellar wind (e.g. \cite{bietenholz21} and references therein).

\section{H.E.S.S. Target of Opportunity program on cc-SNe}

Based on the above-mentioned calculations from Tatischeff \cite{tatischeff09} and Cristofari et al. \cite{cristofari20}, a \hess\ Target of Opportunity (ToO) program has been set up towards nearby cc-SNe in order to probe early VHE $\gamma$-ray emission. Since progenitor mass-loss rates and wind velocities do greatly vary among the different cc-SN sub-classes, with no univocal relationship with the stellar progenitor and the SN type (see e.g. \cite{smartt09b}), the most representative values have been considered as follows: ($\dot{M}_{-5}/u_{\rm w,10}$) = (1,2,2,5,20) for type (II-P,Ib/c,II-L,IIb,IIn) SNe, respectively \cite{nymark06,mauron11,kiewe12,taddia13,smith14}. By assuming the same $\gamma-\gamma$ opacity and CR acceleration efficiency as in SN~1993J, the corresponding horizons of detectability at 5~$\sigma$ with \hess, at day 10 after the SN explosion and for $T_{{\rm obs}}$~=~10~h, have been derived to be (1,1,1,2,8)~Mpc\footnote{The first three values, originally estimated to amount to~0.4-0.8 Mpc, have been rounded to 1 Mpc, below which there is no major star-forming galaxy visible from the Southern Hemisphere except the LMC/SMC.}. Owing to the uncertainties in defining meaningful trigger criteria as described in the previous section, (lower-rank) \hess~ToO observations may also be triggered towards any cc-SN or SN candidate occurring at less than 10~Mpc if there is some interesting MWL information. Given a predicted rate of 1$-$2 cc-SNe per year at $<$ 10 Mpc \cite{smartt09a}, the expected trigger rate does certainly not exceed 2 yr$^{-1}$. This on-going ToO program has been enriched by the implementation of several partnerships with groups dealing with MWL observations of cc-SNe. In particular, one of us (S. Ryder) has been leading a radio ToO program with ATCA at 1-20 GHz towards nearby ($d \lesssim 40$~Mpc), southern (Dec. $<$ -40$^{\circ}$) cc-SNe. Since 2020, several members of the DLT40 and ASAS-SN optical surveys are also involved in this \hess~ToO program. Thus, the \hess~Collaboration can get informed about the properties of any nearby cc-SN event within a few days in order to make a reliable trigger decision.



So far, H.E.S.S. has serendipitously observed nine nearby cc-SNe within a year after their discovery date and triggered ToO observations on one cc-SN a few days after discovery: SN~2016adj, a (possibly compact)\footnote{Type IIb SNe are thought to have more compact and less luminous stellar progenitor with faster and lower-density wind \cite{chevalier10}.} type IIb SN in Cen~A. No significant excess has been found towards any of these cc-SNe.  The analysis and interpretation of the H.E.S.S. data towards these cc-SNe have been published \cite{hess19}. That study provides interesting upper limits on the CSM density under several assumptions on the acceleration efficiency (among other parameters as described in the case of SN~1993J), in particular on SN~2016adj, even though most of the data set spreads over timescales longer than what is considered here. 
In the following these results will be complemented with ToO observations of AT2019krl, originally classified as a type IIn SN candidate, in the galaxy M74.
 
\section{AT2019krl in M74}

\subsection{Discovery and multi-wavelength observations}

On 6th July 2019 a transient originally named ZTF19abehwhj was discovered by ZTF (later renamed as AT2019krl) in the M74 galaxy at $\sim$9.8 Mpc \cite{mcquinn17}. On 9th July AT2019krl was shown to have a type IIn SN optical spectrum, exhibiting narrow and intermediate H-$\alpha$ emission lines, although several types of CSM-interacting SNe, a Luminous Blue Variable (LBV) in outburst, or other intermediate-luminosity transients could not be ruled out at that stage (ATel $\#$12913). Following these reports, a \hess~ToO was triggered on this (possibly very young) type IIn SN candidate on 11th July (see details below). Then, on 12th July, the analysis of the Spitzer archival data to look at the mid-IR pre-discovery evolution of the transient revealed a quiescent source at the location of AT2019krl showing a moderate brightening between December 2018 and April 2019, while the transient became a very bright mid-IR source in mid-May 2019, i.e.~about 50 days before the transient has been discovered by ZTF (ATel $\#$12934). The mid-IR brightness measured at this last epoch suggested that AT2019krl is rather a SN than a LBV outburst, whose final explosion likely occurred between 21st April and 17th May.

\begin{figure*}[ht]%
     {\includegraphics[width=1.00\textwidth,left]{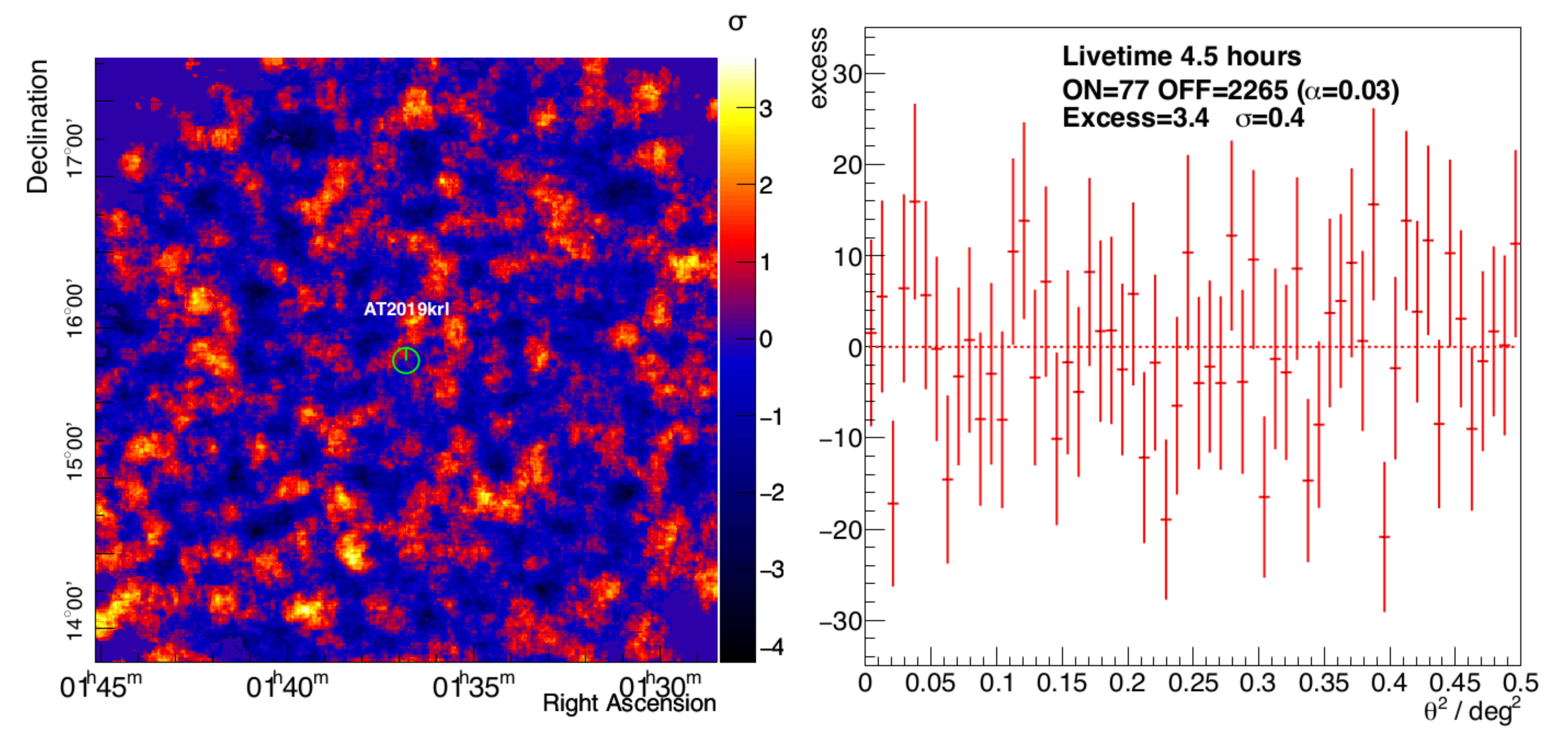}}
    \caption{Significance map (left) and radial distribution of the excess at the source position, in $\theta^{2}$ as described in \cite{Aharonian2006} (right). The reported statistics are explained in the text. The significance of the emission is 0.4 $\sigma$.}
    \label{fig:Sig_Thetasq}
\end{figure*}

\subsection{Observations and data analysis}

Observations were conducted with the High Energy Stereoscopic System (H.E.S.S.), an array of five imaging atmospheric Cherenkov telescopes (IACTs) located in the Khomas Highland of Namibia at an altitude of 1800\,m above sea level. AT2019krl was observed with the full array, the four 12\,m-diameter telescopes (CT1-4, operating since December 2003) and the fifth 28\,m-diameter telescope (CT5, operating since September 2012). The analysis performances are thus the ones of the "H.E.S.S.-II-phase" as described in \cite{Holler_2016}. The data were taken in an "hybrid mode", selecting events triggering at least 3 telescopes including CT5, which allows for a lower energy threshold of 180\,GeV compared to 210\,GeV for the CT1-4 array. 
AT2019krl was observed on three consecutive nights, from 11th to 13th July 2019. These observations were made in wobble mode \citep{Aharonian2006} with a mean offset angle of 0.5 deg around the source and organized in 28-minute exposures called runs. A total of 10 calibrated runs passing the standard quality criteria \citep[see][]{Hahn2014} were analysed using the ImPACT reconstruction \citep{ImPACT}, with hybrid configuration and standard cuts. This led to 4.5 hours of livetime with a mean zenith angle of 47.2 deg. 
The background for the sky map was obtained from rings around each sky bin, applying the ring background method  \cite{Berge2007}.
For the spectral extraction an ON-region (with a radius of 0.086 deg) was defined at the SN position, and multiple OFF regions were selected with the same angular distance from the camera centre as the ON region, using the reflected background method \cite{Berge2007}. The $\gamma$-ray excess was computed using $N_\mathrm{excess} = N_\mathrm{on} - \alpha N_\mathrm{off}$, with $\alpha$ the ON to OFF exposure ratio. The statistical significance was estimated using \citep[][eq. 17]{Li1983}. 
The results were confirmed by an independent data calibration and analysis chain using the Model++ framework \citep{deNaurois2009} with standard quality cuts and the same reflected background method.

\subsection{Results}

After 4.5 h of observations no significant excess has been found. As shown in Figure \ref{fig:Sig_Thetasq}, the significance map is isotropic, and the photon counts as a function of the $\theta^{2}$ distance to the source exhibit no excess above the background level. Flux upper limits (ULs) have thus been derived at the 95$\%$ confidence level under the assumption of a power law spectrum ($dN/dE \propto E^{-\Gamma}$) with a photon index $\Gamma=2$ and 2.4. Integrated ULs at $E > 1$~TeV have been computed in order to compare with ULs previously derived towards cc-SNe \cite{hess19} using a log-likelihood approach \cite{Rolke}. For $\Gamma=2$ [2.4], UL($E > 1$~TeV) = 3.85 [2.95] $\times10^{-13}~\mathrm{cm^{-2}~s^{-1}}$. Assuming a distance to M74 of 9.8 Mpc \cite{mcquinn17}, these flux ULs above 1 TeV translate into luminosity ULs of 3.25 [1.56] $\times 10^{40}$ erg s$^{-1}$.

\section{Discussion and Conclusion}

\begin{figure}[ht]
\begin{minipage}{0.58\textwidth}
    \includegraphics[width=\textwidth]{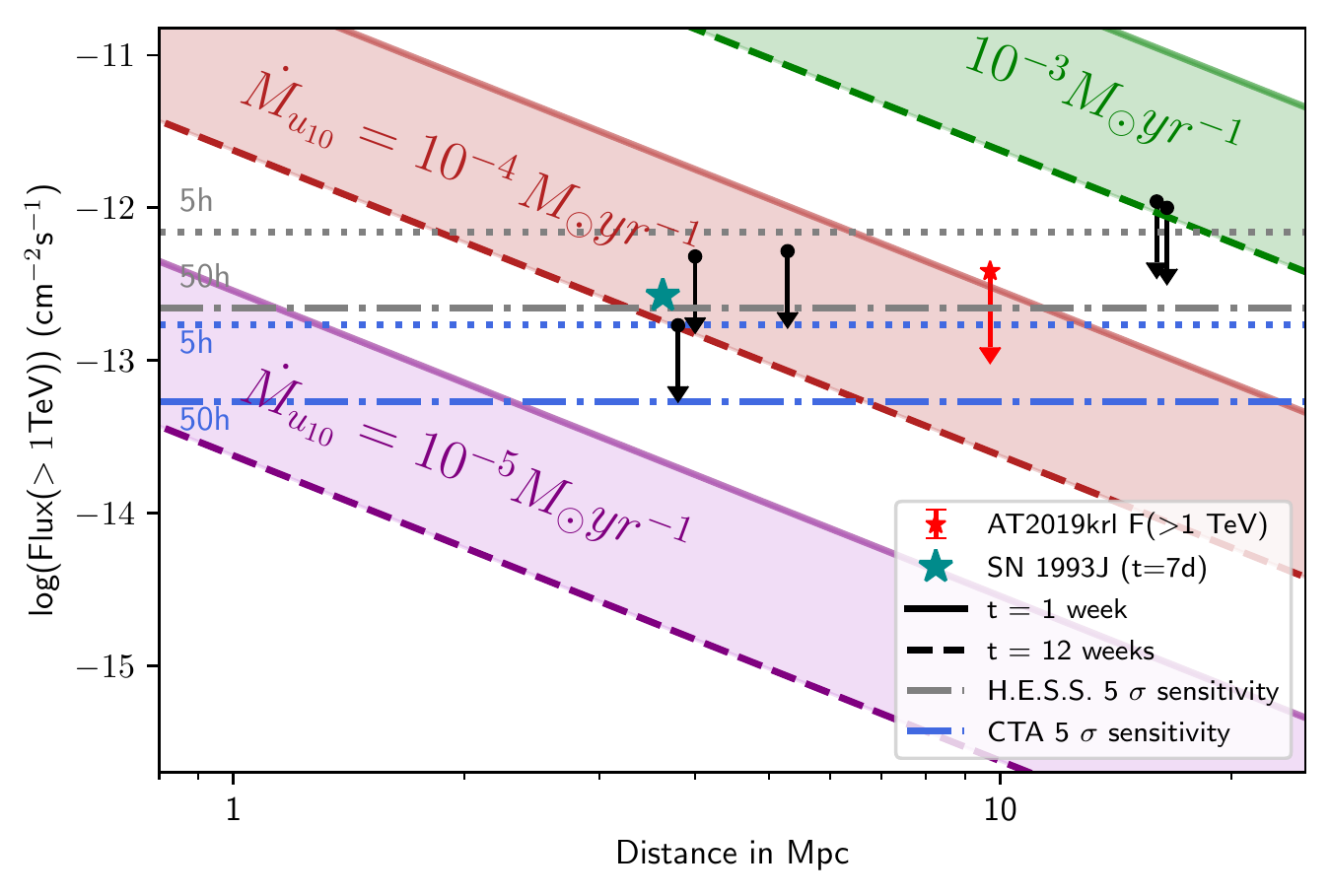} 
\end{minipage}
\begin{minipage}{0.39\textwidth}
  \caption{Predicted flux above 1 TeV as a function of the distance to the source based on \cite{tatischeff09}. $u_\mathrm{w}$ is set to 10~km s$^{-1}$. Fluxes are computed for $t$ = 1 ({solid} lines) and 12 (dashed lines) weeks after the SN explosion. The expected flux from SN 1993J is computed assuming $\dot{M}_{-5}$ = 3.5 \cite{tatischeff09} and $t=7$ days. \hess\ (CT1-4) and CTA integrated sensitivities for 50 h long observations are taken from \cite{Aharonian2006} and \cite{CTA_2019}, from which those for t= 5 h are derived. In black, ULs on the flux above 1 TeV derived in \cite{hess19} are also shown (see text).
  \label{fig:FluxvsDist}}  
\end{minipage}
\end{figure}

Figure~\ref{fig:FluxvsDist} shows that the upper limit on the VHE $\gamma$-ray flux from AT2019krl is at a comparable level as the other cc-SNe studied in \cite{hess19}. The absence of significant $\gamma$-ray emission from AT2019krl suggests in a hadronic model a low density of the target material. The target density can be estimated with the model of \cite{tatischeff09} and using equation (\ref{eq:1}) 
assuming a distance of d = 9.8 Mpc and no pair-production effect. Considering t = 1 week and $\Gamma$=2, the UL on the CSM density ($\dot{M}/u_\mathrm{w}$) amounts to $\sim$6.3 $\times$ 10$^{15}$ g cm$^{-1}$, equivalent to a progenitor mass-loss rate of $\sim 10^{-4}$ (10$^{-3}$) $M_{\odot}$ yr$^{-1}$ for $u_\mathrm{w}$ = 10~(100) km s$^{-1}$, as shown in Fig.~\ref{fig:FluxvsDist}. Such a constraint is at the same level as the lowest wind densities required to make a type IIn SN \cite{smith17}, knowing that most of the well-studied type IIn events exhibit typical mass-loss rates higher by a factor of $\sim$10 \cite{smith17}, for $u_\mathrm{w}$ between 10 and 1000~km s$^{-1}$. Since the absence of $\gamma$-ray emission suggest a low density wind, AT2019krl must either be an atypical SN~IIn, or point to a problem with the $\gamma$-ray emission model. For example, the absence of detectable $\gamma$-ray emission can also indicate a strong attenuation due to the pair-production process with UV/optical photons from the SN photosphere, which is particularly important during the first week after the explosion. Bearing in mind that the date of the explosion of AT2019krl is poorly constrained, if the SN occurred on 21st April (i.e.~t = 12 weeks before the \hess~ToO) as suggested by the analysis of Spitzer archival data (ATel $\#$12934), the UL on the CSM density increases to $\sim$2.2 $\times$ 10$^{16}$ g cm$^{-1}$ but still falls in the lower part of the estimated values for known type IIn SNe \cite[see Fig. 3 in][]{smith17}. Finally, AT2019krl may not have been a SN explosion at all. Indeed, based on extensive archival HST, Spitzer and LBT observations, it has recently been suggested that AT2019krl is either a SN 2008S-like transient due to a Blue SuperGiant (BSG) eruption or an LBV outburst \cite{andrews20}. In this case, the model discussed here cannot be applied. Irrespective of the non-detection of $\gamma$-ray emission and the disputed nature of AT2019krl the study presented here shows that $\gamma$-ray emission from a type IIn SN at a distance of about 10\,Mpc could be detectable with current Imaging Air Cherenkov Telescopes like \hess, provided that the nuclear interaction losses in such high-density CSM are not limiting the acceleration of PeV CRs \citep[see][for detailed calculation]{marcowith18}. The observation of such emission would be a clear proof of efficient particle acceleration during the early stages of a cc-SN, once the progenitor's parameters are known thanks to multi-wavelength observations. 

Estimating the date of the SN explosion is necessary to chose the optimal time window to trigger VHE $\gamma$-ray observations. While shortly after the explosion the $\gamma$-ray emission is absorbed, several weeks later the emission may have declined to a level undetectable by current instruments. The type of the SN explosion is important to estimate the horizon of detectability with IACTs in the TeV domain. Only type IIn SNe can provide a sufficiently dense CSM to produce detectable $\gamma$-ray emission at a distance of about 10\,Mpc while the other types could be detectable at $\sim$1\,Mpc. The detection of a good candidate might necessitate observations longer than considered here, of a few tens of hours. The \hess\ ToO program on cc-SNe is ongoing, allowing for a few ToO observations to be triggered each year. A non-detection by H.E.S.S. of a nearby cc-SNe can provide important constraints on the ability of such very young SNe to accelerate CRs to PeV energies. The future Cherenkov Telescope Array (CTA, \cite{CTA_2019}), with a higher sensitivity than \hess, will be able to probe the VHE $\gamma$-ray emission from cc-SNe at a much lower level.

\small{
\paragraph{Acknowledgements}

The support of the Namibian authorities and of the University of Namibia in facilitating the construction and operation of H.E.S.S. is gratefully acknowledged, as is the support by the German Ministry for Education and Research (BMBF), the Max Planck Society, the German Research Foundation (DFG), the Helmholtz Association, the Alexander von Humboldt Foundation, the French Ministry of Higher Education, Research and Innovation, the Centre National de la Recherche Scientifique (CNRS/IN2P3 and CNRS/INSU), the Commissariat à l'énergie atomique et aux énergies alternatives (CEA), the U.K. Science and Technology Facilities Council (STFC), the Knut and Alice Wallenberg Foundation, the National Science Centre, Poland grant no. 2016/22/M/ST9/00382, the South African Department of Science and Technology and National Research Foundation, the University of Namibia, the National Commission on Research, Science \& Technology of Namibia (NCRST), the Austrian Federal Ministry of Education, Science and Research and the Austrian Science Fund (FWF), the Australian Research Council (ARC), the Japan Society for the Promotion of Science and by the University of Amsterdam. We appreciate the excellent work of the technical support staff in Berlin, Zeuthen, Heidelberg, Palaiseau, Paris, Saclay, T\"{u}bingen and in Namibia in the construction and operation of the equipment. This work benefitted from services provided by the H.E.S.S. Virtual Organisation, supported by the national resource providers of the EGI Federation.}

{

}


\clearpage
\section*{Full Authors List: \Coll\ Collaboration}
\scriptsize
\noindent
H.~Abdalla$^{1}$, 
F.~Aharonian$^{2,3,4}$, 
F.~Ait~Benkhali$^{3}$, 
E.O.~Ang\"uner$^{5}$, 
C.~Arcaro$^{6}$, 
C.~Armand$^{7}$, 
T.~Armstrong$^{8}$, 
H.~Ashkar$^{9}$, 
M.~Backes$^{1,6}$, 
V.~Baghmanyan$^{10}$
% ...

\medskip

\noindent
$^{1}$University of Namibia, Department of Physics, Private Bag 13301, Windhoek 10005, Namibia\\
$^{2}$Dublin Institute for Advanced Studies, 31 Fitzwilliam Place, Dublin 2, Ireland\\
$^{3}$Max-Planck-Institut f\"ur Kernphysik, P.O. Box 103980, D 69029 Heidelberg, Germany\\
$^{4}$High Energy Astrophysics Laboratory, RAU,  123 Hovsep Emin St  Yerevan 0051, Armenia\\
$^{5}$Aix Marseille Universit\'e, CNRS/IN2P3, CPPM, Marseille, France\\
$^{6}$Centre for Space Research, North-West University, Potchefstroom 2520, South Africa\\
$^{7}$Laboratoire d'Annecy de Physique des Particules, Univ. Grenoble Alpes, Univ. Savoie Mont Blanc, CNRS, LAPP, 74000 Annecy, France\\
$^{8}$University of Oxford, Department of Physics, Denys Wilkinson Building, Keble Road, Oxford OX1 3RH, UK\\
$^{9}$IRFU, CEA, Universit\'e Paris-Saclay, F-91191 Gif-sur-Yvette, France\\
$^{10}$Instytut Fizyki J\c{a}drowej PAN, ul. Radzikowskiego 152, 31-342 Krak{\'o}w, Poland\\
% ...

%
%
%

\end{document}